Negative magnetoresistance dynamics in expanded graphite under hydrostatic pressure up to 1.8 GPa.

V.V. Slyusarev, P.I. Polyakov
Institute for Physics of Mining Processes, N.A.S. Ukraine, R. Luxemburg 72, Donetsk 83114, Ukraine.

Basal plane resistivity of expanded graphite was studied under simultaneous influence of hydrostatic pressure up to 1.8 GPa and magnetic field 0.8 T in the 77-300 K temperature region. Magnetic field induces negative magnetoresistance in the sample within all temperature and pressure range studied. A change in resistivity of the sample under maximum pressure reaches 80%. Significant change in resistivity dependence on temperature under the pressure of 0.6 GPa suggests for ordering transition in the sample studied. Negative magnetoresistance in the graphite reaches about 15% at 0.6 GPa. Magnetic field acts in the same way as pressure and potentiates the transition formation and further magnetoresistance dynamics. The effects observed are mostly of elastic character according to resistivity of the unloaded sample.

## 1. Introduction

Rising interest on magnetism in carbon structures widely described in literature is called foremost by the cheapness and availability of this material which unique combination of properties opens a rich feasibility in wide range of practical use. Graphene invention inspired inter alia by the interest in intercalated and plane graphite brought the interest back to the latter [1]. The magnetic field effect on the conductivity of graphite single crystals with a change from a square form of dependence to the line one and the significant influence of the impurities and imperfections in the crystal was shown as far as in [2]. Magnetism in graphites is mainly connected with the presence of edge-states, huge number of unbounded atoms, defects, surfaces and pores where size effects and mesoscopic structure are of significant role [3-5]. Thus, negative magnetoresistance was described for soft carbons and graphite under different thermal treatment and for artificial and kish-graphite under its structure disordering [6, 7]. Anisotropy, a 30-times compressibility difference in the parallel and perpendicular direction to the c-axe leads to the fact, that the volume's compressibility is determined mainly by c-axe compressibility, which is the effect of weak interactions between the planes [8, 9]. Thereby, the influence of hydrostatic pressure on weak electrostatic forces in distorted structure may have a more significant effect. High pressure does not lead to structural transitions in nanotubes [10] till unzipping [11], but transitions have been observed in fullerenes [12] and plane structures [13-15], being associated mainly with ordering. In general, results of the structure analysis of graphite's under pressure are more numerous than experimental results on properties, better studied under magnetic field influence. Simultaneous effect of high pressure, magnetic field and temperature on resistivity of expanded graphite presented in this study may be of multidisciplinary interest.

## 2. Experimental

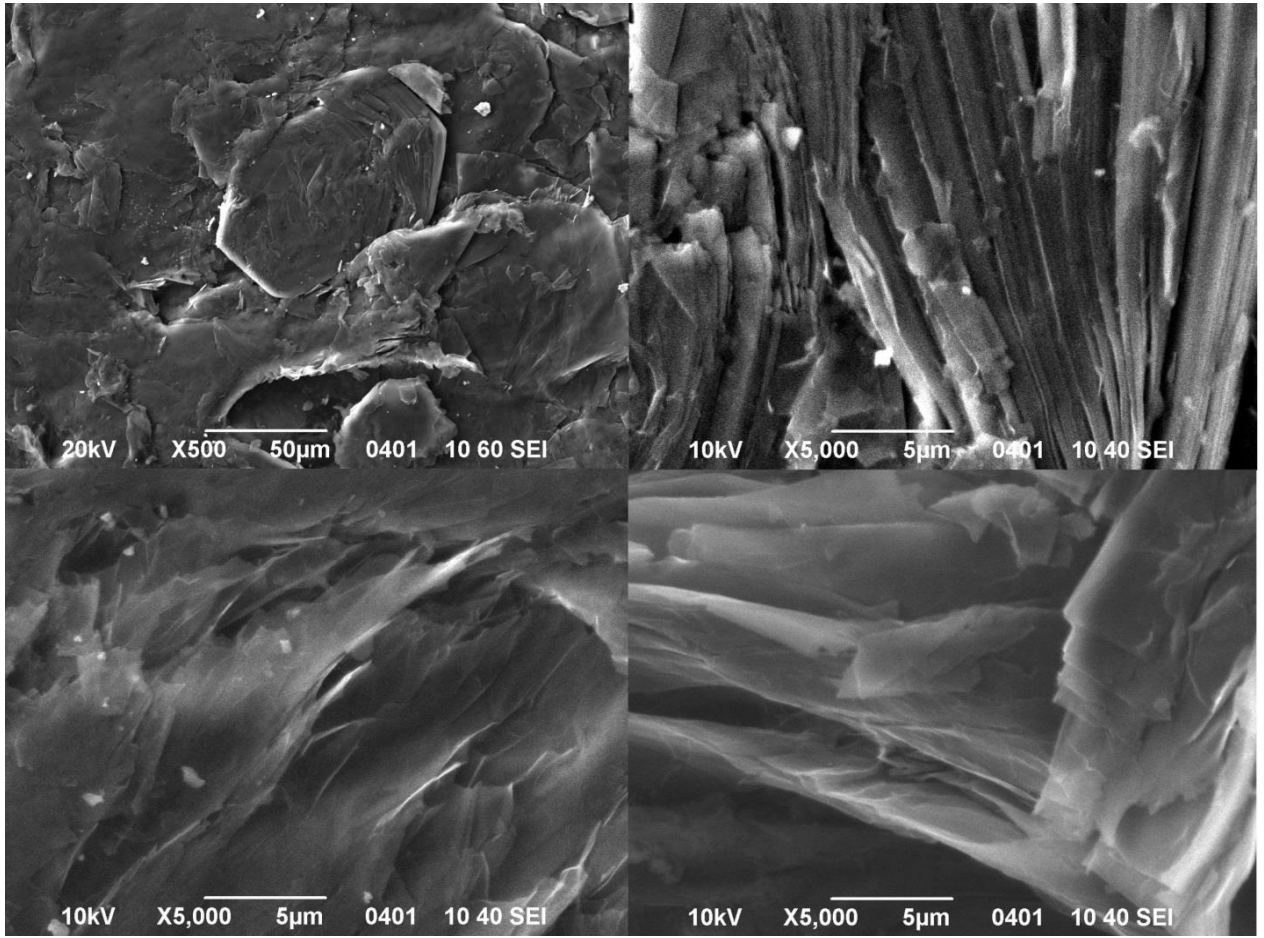

Fig. 1 –Structure of the spacemen studied.

Thermoexfoliated graphite with 1.91 g/cm$^3$ density was studied. Sample represented by L.Y. Matzui group from Kiev National University [16-18]. Formed by hot isostatic pressing to dimensions 7×3×1.6 mm the sample contains interparticle and innerparticle porosity and consists of hundred nanometers domains packed in typical structure, shown on Fig 1. Only tiny amount of sulfur has been detected in the sample which is due to natural precursors. Electric contacts were made by silver paste. Measurements performed in basal plane by a four-probe method. To prevent the interaction between liquid media in the high pressure cell and the porous sample, the latter, with mounted electrodes, was coated by thin wax film. No breakage of the wax film was estimated after pressure relief. To obtain the hydrostatic pressure [19] up to 1.8 GPa double-layered nonmagnetic cell was used shown on fig. 2 and in detail described with methodology in [20]. Measurements performed under fixed pressures in 77 – 300 K temperature range in zero magnetic field and weak magnetic field 0.8 T, applied perpendicular to the basal plane. A copper resistance thermometer inside the cell was used for temperature measurement with precision of 0.25 K. Pressure value was estimated with a precision around 50 MPa by manganine gauge. Resistivity values have been calculated from specimen geometry as voltage on potential electrodes at direct current 270 mA.

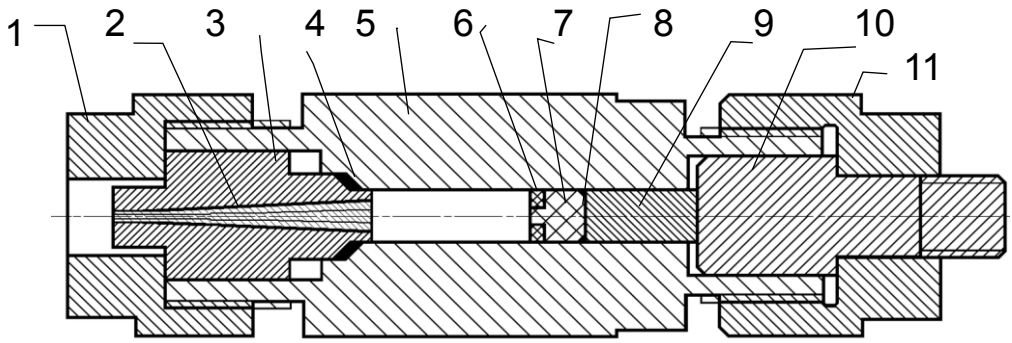

Fig. 2 - High pressure cell (1- obturator check-nut; 2- electric wires; 3- obturator; 4- obturator seal; 5- double-layered body; 6- rubber ring; 7- ptfs seal; 8- bronze ring; 9- piston; 10- plunger; 11- check nut).

## 3. Results

Experimental data show more times higher pressure effect on the sample resistance than temperature and magnetic field effect (fig. 3). Resistivity change reaches an 80% rate in the studied pressure range (insert on fig. 3), which in general suggests with results for these similar structures [13-15]. Experimental dots for the unloaded sample show quite good coincidence with initial dependence of resistivity, which commonly attributes to hydrostatic conditions and elastic character of pressure effect on porous structure of the sample in experimental procedure.

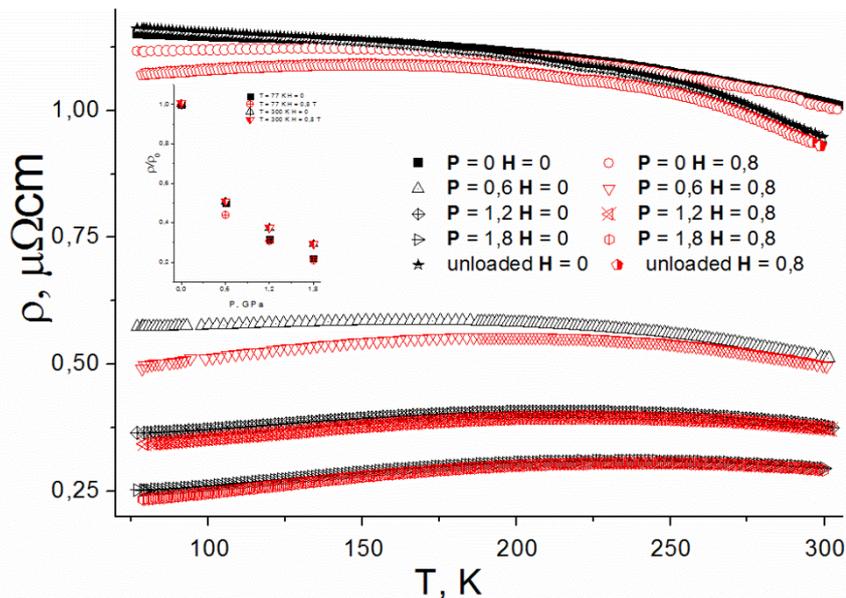

Fig. 3 – resistivity dependence on temperature at fixed pressure P, GPa and magnetic field H, T (relative resistivity change under pressure on insert).

Consecutive consideration of the resistivity dependences at fixed pressures (fig. 4) shows that under the normal pressure in the studied temperature range a typical semimetal form of the resistivity dependence on temperature is observed for the sample studied. Applying of magnetic field (0.8 T) leads to negative magnetoresistance appearance in the low temperatures range with a little maxima (Fig.4, a). Application of the pressure of 0.6 GPa leads to maxima appearance in zero field and "metallic" form of the dependence occurs in the magnetic field (Fig. 4, b). Increasing the pressure to 1,2 GPa leads to "metallic" part formation in zero field and shifting of the maxima to higher temperatures at 0.8 T (Fig. 4, c). Further increasing of the pressure to 1.8

GPa leads to the maxima shifting in higher temperature range and negative magnetoresistance suppressing (Fig. 4, d).

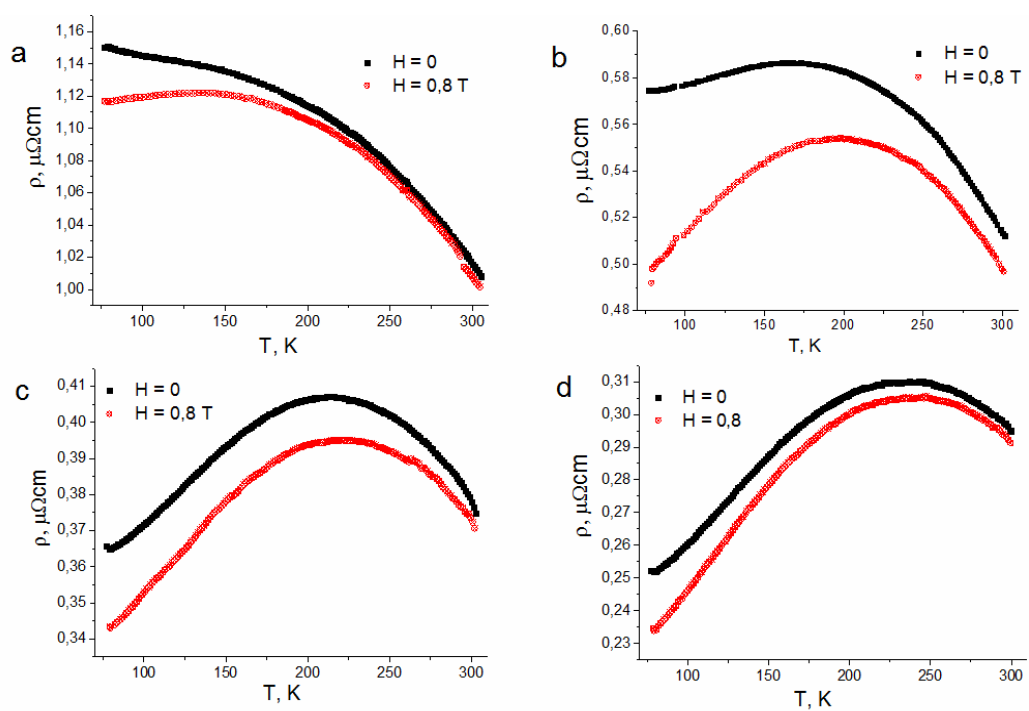

Fig. 4 – Resistivity at fixed pressures (a – 0 GPa, b – 0.6 GPa, c – 1.2 GPa, d – 1.8 GPa).

Normalized by the maxima ($\rho/\rho_{max}$), relative resistivity dependence under pressure and magnetic field (fig. 5) shows a complicated character which points to different qualitative resistivity changes in left and right parts of the dependence relative to the maxima. Application of the pressure and magnetic field leads to relative resistivity decrease in "metallic" left part from the low temperatures region while in the right semimetal part from high temperatures range relative resistivity increases.

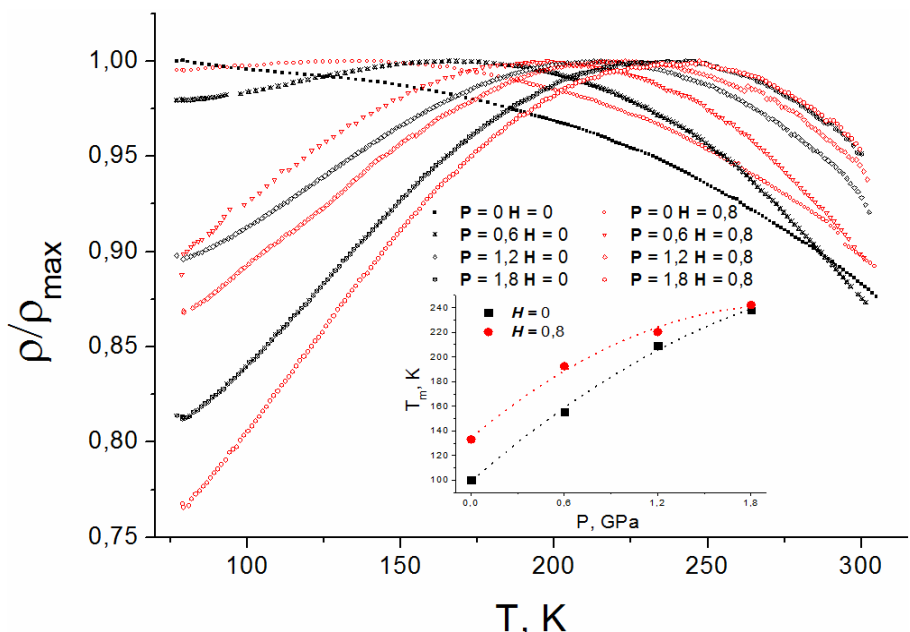

Fig. 5 – relative resistivity at fixed pressure P, GPa and magnetic field H, T (maxima temperature shift under pressure on insert).

Insert on fig. 5 shows temperature of the maxima shift under pressure. Position of the maxima was calculated out of second order polynomial function derivative. Character of the resistivity maxima temperature dependence on pressure suggests for possibility of magnetoresistance sign change at relatively higher pressures with increasing of sample density. Magnetoresistance under pressure (shown on fig. 6) has clear minimum at almost all temperatures due to the non-synchronism of the transition in magnetic field, but its relative value on temperature (MR/MR$_{77}$) weakly affected by pressure.

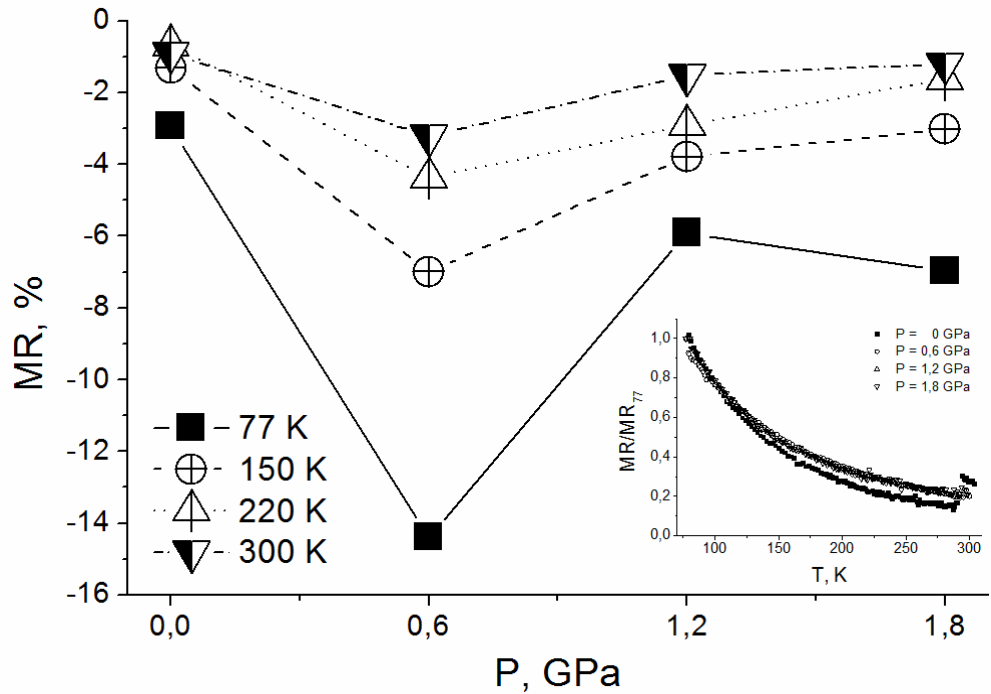

Fig. 6 – magnetoresistance under pressure at various temperatures (relative magnetoresistance on temperature under fixed pressure on insert).

## 4. Discussion

Negative coefficient of resistivity in microfine graphite structures is usually attributed to weak localization and interaction effects, preferentially associated with the current carrier scattering on crystallite boundaries and resistivity maxima shifting suggests for the carriers mean free path change [3]. Thus, resistivity of the graphite, determined by in-plane strong covalent bonding, affected by pressure more than it may be expected, because between-the-plane resistivity due to the structure anisotropy plays in order of magnitude less significant role [21].

Composition of the sample demonstrated in fig. 1 and expanded graphite structure evolution under formation to monolith described for relatively low pressures in [22-24] does not only suggest the presence of huge amount of pores, edges and surfaces but also infers their leading role in macro- and mesoscopic structure formation. The structure in the whole becomes cross-linked not by strong covalent bonds but weak Van der Waals bonds, and the latter determining flexible properties of such kind of graphite. Pressure effect converges to influence on this energetically comparable [20] with it weak bonds and through them on sample resistivity value in whole. Such not quite dense structure enables a possibility of hydrostaticity loss inside the sample due to the presence of stress concentrators which may lead to microplastic deformation. Presence of shear stresses on the one hand leads to additional sample densification but on the other – it increases the number of defects and edges. This may be the reason of negative magnetoresistivity effect increasing and the resistivity dependence slope sharpening in high temperature region for the unloaded sample.

Experimental data suggests for the structure transition under pressure about 0.6 GPa in the sample studied. Resistivity change of about an order of magnitude may be explained not only by ordering transition itself but a with a percolation threshold too. Furthermore, turbostratic and porous structure of the sample enables a possibility of the expanded graphite domains bending and additional stresses generation directly in the planes, as it shown for pore coalescence anisotropy [24] and non-uniform strain distribution [23] under the relatively lower pressures, thus the order of magnitude higher pressures will affect more fine elements more intensively. Induced stresses in the graphene plane may lead to the effects similar to magnetic field effects [24]. Additional stresses with values exceeding external pressure values, according to Hertz problem, may arise directly on the contacts of the constituent parts and the stresses value strongly depends of contact geometry. A possibility of bending stresses appearance in expanded graphite structure has been illustrated on right bottom microphotograph of fig. 1, where an element of half-pipe shape may be seen. A possible explanation for this element may lay in the thermodynamic conditions of the sample formation, when in some particular regions a premises to form a carbon nanotube are created, but the massive domains connected by twins in "wormlike" particles prevents it under high pressing load.

A question may be asked about the possibility to create a material with properties determined not by the internal structure itself but by the properties of borders, surfaces and defects through the specific sizes and distances, which local fields will shape a total electromagnetic field in the sample. These properties tighten with Van der Waals interaction will be strongly affected by pressure and magnetic field as ruling parameter even at relatively low values. And this may be practically realized for nanostructures in shrinking coverage.

Influence of hydrostatic pressure principally differs from the axial nonquecomponent one due to the absence of shear stresses components on the border of the sample and uniform all-around compression in liquid, what in these conditions may generate a local inhomogeneous stress fields within the turbostratic structure of the sample. In the pressure region about 0.6 GPa, due to the transition formation and non-equilibrium state occurring, with no possibility of relaxation by shear in basal plane, this stresses probably leads to inhomogeneous electromagnetic field formation, and application of the weak external magnetic field is enough to converse the system to another state.

**4. Conclusions**

Simultaneous influence of pressure and magnetic field on the sample resistivity in wide temperature region represents a powerful tool for its structure-properties relations investigation. Magnetic field induces negative magnetoresistance in nanostructured expanded graphite sample within all temperatures and pressures range studied. A change in resistivity of the sample under maximum pressure reaches 80%. Significant change in resistivity dependence on temperature under the pressure of 0.6 GPa suggests for ordering transition in the sample studied. Negative magnetoresistance reaches about 15% due to the non-synchronism of the transition. Magnetic field affects resistivity in the same way as pressure, potentiating the transition formation and further magnetoresistance dynamics in the sample studied. The effects observed are mostly of elastic character according to unloaded sample resistivity dependence.